# Tin-DNA Complexes Investigated by Nuclear Inelastic Scattering of Synchrotron Radiation

**G. Barone**
*Dipartimento di Chimica Inorganica e Analitica, Universitá degli Studi di Palermo, I-90128 Palermo, Italy*

**L. H. Böttger, J. A. Wolny, H. Paulsen and A. X. Trautwein**
*Institut für Physik, Universität zu Lübeck, D-23538, Lübeck, Germany*

**A. Silvestri**
*Dipartimento di Chimica Inorganica e Analitica, Universitá degli Studi di Palermo, I-90128 Palermo, Italy*

**I. Sergueev**
*ESRF, F-38043 Grenoble Cedex 9, France*

**G. LaManna**
*Dipartimento di Chimica Fisica, Universitá degli Studi di Palermo, I-90128 Palermo, Italy*

**Abstract.** Nuclear inelastic scattering (NIS) of synchrotron radiation has been used to investigate the dynamics of tin ions chelated by DNA. Theoretical NIS spectra have been simulated with the help of density functional theory (DFT) calculations using twelve models for different binding sites of the tin ion in $(CH_3)Sn(DNA_{Phosphate})_2$. The simulated spectra are compared with the measured spectrum of the tin-DNA complex.

**Keywords:** nuclear inelastic scattering, tin-DNA

**Abbreviations:** NIS - nuclear inelastic scattering; DFT - density functional theory; PVDOS - partial vibrational density of states

## 1. Introduction

Organotin(IV) compounds are generally toxic and due to chemical applications of their derivatives, widely spread in the environment. A possible explanation of their toxicity seems to be their ability to interact with DNA. The interactions between organotin cations and DNA have been studied by [119]Sn-Mössbauer spectroscopy and by electronic structure calculations [1]. The Mössbauer studies suggest a direct octahedral coordination of the organotin cation to the DNA, most probably to the negatively charged phosphate groups. Electronic structure calculations at the level of the semiempirical method PM3 [2, 3], have been performed for the determination of geometries. PM3 is especially suitable to predict intermolecular hydrogen bondings. The geometries obtained are generally in good agreement with [119]Sn Mössbauer and X-ray data. Calculations have also been performed for a model complex of dinucleoside triphosphate duplex (pApCp)-(pTpGp), which mimics DNA, and the cationic dimethyltin(IV) moieties in different binding positions. Among the possible DNA coordination sites for metal ions only the two most important are mentioned here i.e., the oxygen atoms of phosphate groups and the heteroatoms of nitrogen bases. Four structures have been obtained (DD1 to DD4) by calculations for the case of interaction with phosphate groups and for the case of interaction with nitrogen bases, respectively [4]. Electronic structure calculations also provide information about the dynamical behaviour of the $^{119}$Sn nucleus. The dynamics are strongly connected with the strength of the bonds between tin and its organic ligands. Experimental access to the dynamics via conventional vibrational spectroscopy (IR, Raman) is hampered by the complexity of the spectra which are dominated by the vibrations of the large organic ligands. Therefore, our aim was to focus on those vibrational modes connected with a significant contribution of the tin nucleus. For this reason we have applied nuclear inelastic scattering of synchrotron radiation (NIS) to detect tin-centered molecular vibrations. The measured NIS spectra allow for the first time to determine the frequency of bond stretching vibrations of tin atoms bound to DNA fragments. The measured spectra are compared with simulated NIS spectra obtained from electronic structure calculations [5]. This comparison corroborates our hypothesis about the binding sites of tin [6]. To our knowledge, NIS measurements on biological samples (e.g. iron porphyrins or rubredoxines) have been performed up to now only in cases where the structure of the binding site is known by other methods like X-ray diffraction. This study thus demonstrates the usefulness of the NIS method in an extended field of biological samples.

## 2. Experimental

A detailed description of the sample preparation is given in ref. [6]. Measurements were performed at beamline ID22 [7, 8] ESRF Grenoble. The beamline setup was optimized for measurements at the Sn-resonance energy of 23.875 keV. The sample was placed in a slit sample holder (10 mm x 3 mm x 2 mm) and attached to a constant flow cryostat. The temperature was set to 50 K and had a temperature stability of about 1 K. The 10 mm slit of the sample holder was orientated parallel to the synchrotron beam and the detector for the delayed photons was placed directly beneath (~5 mm) the sample and perpendicular to the synchrotron beam, with the detector area covering the whole length of the slit. The energy resolution of the beamline setup was about 0.8 meV. The NIS spectra were collected in multiple scans, each having an energy step width of 0.2 meV.

## 3. Results and Discussion

The NIS spectra were processed by the method of Kohn [9] to calculate the partial vibrational density of states (PVDOS) [10]. After the subtraction of the elastic scattering contribution [10] the NIS spectra show at least 5 shoulders on the positive energy side, which also appear in the phonon density of states as 5 clear peaks (c-g) in the energy range of 11 to 27 meV and as 2 clear shoulders (a-b) between 8 and 11 meV (Fig. 1). The spectral region above 27 meV is not discussed here because of a very low signal to noise ratio.
Barone et al. [4, 6] have considered 12 possible configurations of organotin-DNA complexes in their electronic structure calculation at the level of density functional theory and PM3 respectively. Two models (DD1 and DD2) describe the Sn-O-P interaction with two symmetrically placed water molecules around the tin atom and with two water molecules placed on the same side of the tin atom respectively. Two other models (DD3 and DD4) describe the interaction of the tin atom via different nitrogen-molecules of the DNA. Detailed description of the models can be found in ref. [4, 6].

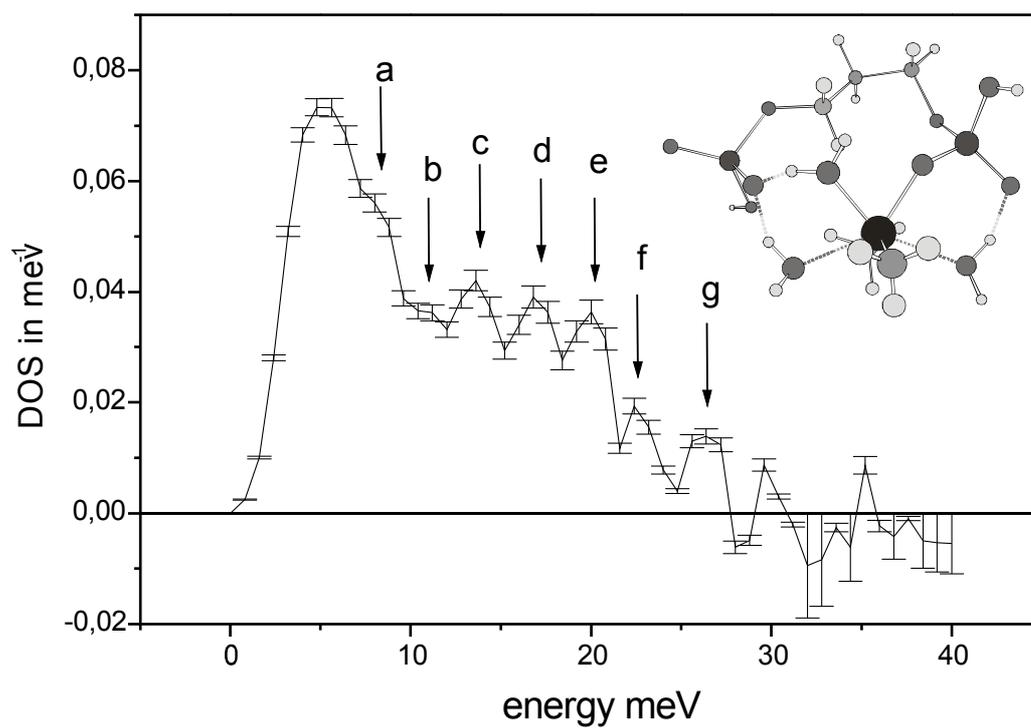

*Figure 1.* PVDOS extracted from the measured NIS spectrum. The inset shows the structure model used for calculation of model DD1.

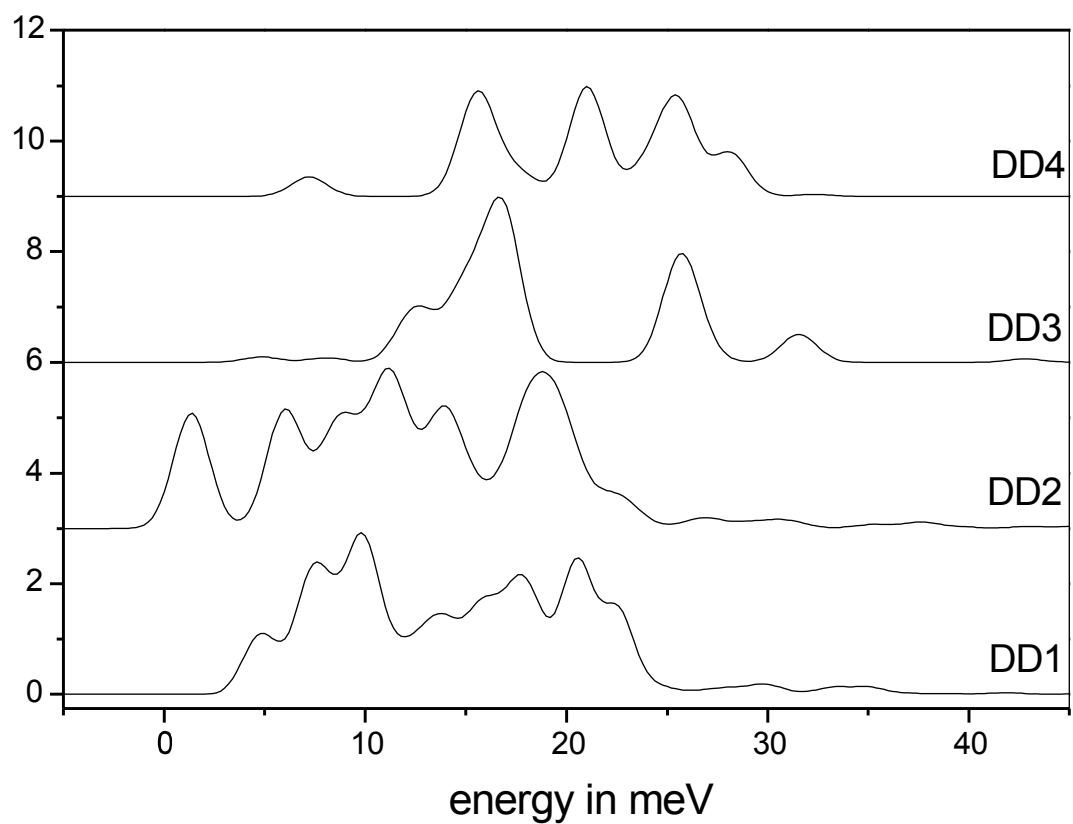

*Figure 2.* Calculated PVDOS based on DFT calculations for models DD1 to DD4.

Models DD3 and DD4 yields in our calculations 3 almost equally strong peaks in the range of 15 meV to 25 meV, but had different position and energy spacing compared to experiment. Model DD2 yields peaks, which are in between observed peak positions. The obtained peak positions for model DD1 resemble closed those of the observed spectra. This model exhibits an additional peak (resolved as shoulder) at about 16 meV in reasonable agreement with peak (d) of the experimental spectrum (Fig 1). The calculated normal modes that we assigned to the observed peaks are strongly mixed. The main contributions are in all cases Sn-OH$_2$ bond stretching and H$_2$O-Sn-OH$_2$ bending modes.

## 4. Conclusion

From comparing calculated PVDOS with PVDOS extracted from a measured NIS spectrum for different tin-ligand configurations we conclude that the tin atom is bound through an oxygen bridge towards the phosphor of the DNA fragment and has symmetrically arranged water molecules around the tin atom. Thus nuclear inelastic scattering together with electronic structure calculations is very suitable to study the metal-ligand coordination geometry and its vibrational behaviour of Mössbauer-active metal centres in even large biomolecules such as DNA.

## Acknowledgements

One of the authors (J.A.W.) gratefully acknowledges financial support by the Deutsche Forschungsgemeinschaft (DFG, Tr 97/32).

## References

1. R. Barbieri, G. Ruisi, A. Silvestri, A. M. Giuliani, A. Barbieri, G. Spina, F. Pieralli and F. Del Giallo, Journal of the Chemical Society, Dalton Transactions: Inorganic Chemistry p. 467.
2. J. J. P. Stewart, Journal of Computational Chemistry 10 (1989) 209.
3. J. J. P. Stewart, Journal of Computational Chemistry 12 (1991) 320.
4. G. Barone, M. C. Ramusino, R. Barbieri and G. La Manna, Journal of Molecular Structure (Theochem) 469 (1999) 143.
5. H. Paulsen, R. Benda, C. Herta, V. Schünemann, A. I. Chumakov, L. Duelund, H. Winkler, H. Toftlund and A. X. Trautwein, Physical Review Letters 86 (2001) 1351.
6. G. Barone, L. H. Böttger, J. A. Wolny, H. Paulsen, A. X. Trautwein, A. Silvestri, I. Sergueev, R. Rüffer and G. LaManna, submitted to Hyperfine Interactions.
7. A. Chumakov and R. Rüffer, Hyperfine Interactions 113 (1998) 59.
8. R. Rüffer and A. I. Chumakov, Hyperfine Interactions 97-98 (1996) 589.
9. V. G. Kohn, A. I. Chumakov and R. Rüffer, Physical Review B 58 (1998) 8437.
10. W. Sturhahn, T. S. Toellner, E. E. Alp, X. Zhang, M. Ando, Y. Yoda, S. Kikuta, M. Seto, C. W. Kimball and B. Dabrowski, Physical Review Letters 74 (1995) 3832.